\documentclass[12pt]{article}
\usepackage{html}
\usepackage{epsf}
\usepackage[pdftex]{graphicx}
\usepackage{graphicx}
\usepackage{amssymb}
\usepackage{amsmath}
\usepackage{hyperref}
\begin{document}
\begin{titlepage}
\includegraphics[width=150mm]{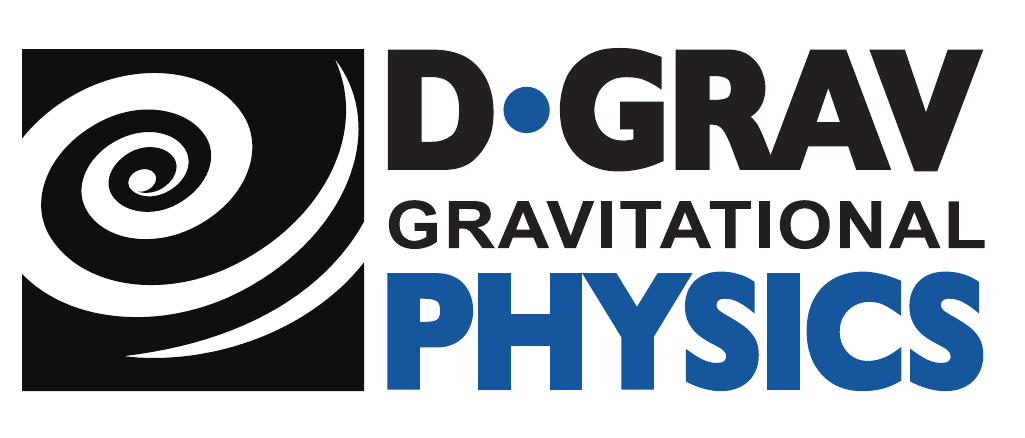}
\begin{center}
{ \Large {\bf MATTERS OF GRAVITY}}\\ 
\bigskip
\hrule
\medskip
{The newsletter of the Division of Gravitational Physics of the American Physical 
Society}\\
\medskip
{\bf Number 54 \hfill December 2019}
\end{center}
\begin{flushleft}
\tableofcontents
\end{flushleft}
\end{titlepage}
\vfill\eject
\begin{flushleft}
\section*{\noindent  Editor\hfill}
David Garfinkle\\
\smallskip
Department of Physics
Oakland University
Rochester, MI 48309\\
Phone: (248) 370-3411\\
Internet: 
\htmladdnormallink{\protect {\tt{garfinkl-at-oakland.edu}}}
{mailto:garfinkl@oakland.edu}\\
WWW: \htmladdnormallink
{\protect {\tt{http://www.oakland.edu/physics/Faculty/david-garfinkle}}}
{http://www.oakland.edu/physics/Faculty/david-garfinkle}\\
\section*{\noindent  Associate Editor\hfill}
Greg Comer\\
\smallskip
Department of Physics and Center for Fluids at All Scales,\\
St. Louis University,
St. Louis, MO 63103\\
Phone: (314) 977-8432\\
Internet:
\htmladdnormallink{\protect {\tt{comergl-at-slu.edu}}}
{mailto:comergl@slu.edu}\\
WWW: \htmladdnormallink{\protect {\tt{http://www.slu.edu/arts-and-sciences/physics/faculty/comer-greg.php}}}
{http://www.slu.edu/arts-and-sciences/physics/faculty/comer-greg.php}\\
\bigskip
\hfill ISSN: 1527-3431


\bigskip

DISCLAIMER: The opinions expressed in the articles of this newsletter represent
the views of the authors and are not necessarily the views of APS.
The articles in this newsletter are not peer reviewed.

\begin{rawhtml}
<P>
<BR><HR><P>
\end{rawhtml}
\end{flushleft}
\pagebreak
\section*{Editorial}

This issue of Matters of Gravity is the last one that I edit.  The next Editor will be Aaron Zimmerman.  We all wish Aaron good luck with his new responsibilities.

The next newsletter is due June 2020.  Issues {\bf 28-54} are available on the web at
\htmladdnormallink 
{\protect {\tt {https://sites.google.com/oakland.edu/garfinkl}}}
{https://sites.google.com/oakland.edu/garfinkl} 
All issues before number {\bf 28} are available at
\htmladdnormallink {\protect {\tt {http://www.phys.lsu.edu/mog}}}
{http://www.phys.lsu.edu/mog}

Any ideas for topics
that should be covered by the newsletter should be emailed to me, or 
Greg Comer, or
the relevant correspondent.  Any comments/questions/complaints
about the newsletter should be emailed to me.

A hardcopy of the newsletter is distributed free of charge to the
members of the APS Division of Gravitational Physics upon request (the
default distribution form is via the web) to the secretary of the
Division.  It is considered a lack of etiquette to ask me to mail
you hard copies of the newsletter unless you have exhausted all your
resources to get your copy otherwise.

\hfill David Garfinkle 

\bigbreak

\vspace{-0.8cm}
\parskip=0pt
\section*{Correspondents of Matters of Gravity}
\begin{itemize}
\setlength{\itemsep}{-5pt}
\setlength{\parsep}{0pt}
\item Daniel Holz: Relativistic Astrophysics,
\item Bei-Lok Hu: Quantum Cosmology and Related Topics
\item Veronika Hubeny: String Theory
\item Pedro Marronetti: News from NSF
\item Luis Lehner: Numerical Relativity
\item Jim Isenberg: Mathematical Relativity
\item Katherine Freese: Cosmology
\item Lee Smolin: Quantum Gravity
\item Cliff Will: Confrontation of Theory with Experiment
\item Peter Bender: Space Experiments
\item Jens Gundlach: Laboratory Experiments
\item Warren Johnson: Resonant Mass Gravitational Wave Detectors
\item David Shoemaker: LIGO 
\item Stan Whitcomb: Gravitational Wave detection
\item Peter Saulson and Jorge Pullin: former editors, correspondents at large.
\end{itemize}
\section*{Division of Gravitational Physics (DGRAV) Authorities}
Chair: Gary Horowitz; Chair-Elect: Nicolas Yunes
; Vice-Chair: Gabriela Gonzalez . 
Secretary-Treasurer: Geoffrey Lovelace; Past Chair: Emanuele Berti ; Councilor: Beverly Berger
Members-at-large:
Lisa Barsotti, Theodore Jacobson, Michael Lam, Jess McIver, Alessandra Corsi, Henriette Elvang.
Student Members: Belinda Cheeseboro, Alejandro Cardenas-Avendano.
\parskip=10pt

\vfill\eject

\section*{\centerline
{we hear that \dots}}
\addtocontents{toc}{\protect\medskip}
\addtocontents{toc}{\bf DGRAV News:}
\addcontentsline{toc}{subsubsection}{
\it we hear that \dots , by David Garfinkle}
\parskip=3pt
\begin{center}
David Garfinkle, Oakland University
\htmladdnormallink{garfinkl-at-oakland.edu}
{mailto:garfinkl@oakland.edu}
\end{center}

Kip Thorne has been awarded the James Madison Medal of Princeton University.

Bernard Schutz and Bruce Allen have been awarded the APS Richard A. Isaacson Award in Gravitational-Wave Science.

Steven Christensen, Matthew Evans, Jocelyn Read, B.S. Sathyaprakash, and Peter Shawhan have been elected APS Fellows.

Hearty Congratulations!

\vfill\eject
\section*{\centerline
{APS April Meeting}}
\addtocontents{toc}{\protect\medskip}
\addcontentsline{toc}{subsubsection}{
\it APS April Meeting, by David Garfinkle}
\parskip=3pt
\begin{center}
David Garfinkle, Oakland University
\htmladdnormallink{garfinkl-at-oakland.edu}
{mailto:garfinkl@oakland.edu}
\end{center}
We have a very exciting DGRAV related program at the upcoming APS meeting April 18-21 in Washington, DC.  Our Chair-elect, Nicolas Yunes, did an excellent job of putting together this program.

\vskip0.2truein
DGRAV related plenary sessions include talks by James Peebles and Michel Mayor, as well as the following:
\vskip0.2truein
Laura Cadonati, {\it LIGO/Virgo: Multi-messenger observations}\\
Amina Helmi, {\it GAIA observations}\\
Zaven Arzoumanian, {\it NICER news}\\
Didier Queloz, {\it Exoplanets}\\
\vskip0.2truein
There will also be a public lecture by Shep Doeleman on the Event Horizon Telescope.

\vskip0.2truein
The DGRAV sponsored invited sessions are\\
\vskip0.2truein
{\bf New results from LIGO} \\
Chad Hanna, {\it Compact Binaries in Advanced LIGO's Third Observing Run}\\
Maya Fishbach, {\it Astrophysical Lessons from LIGO/Virgo's Black Holes}\\
Max Isi, {\it Some Highlights from LIGO and Virgo’s Third Observing Run}\\
\vskip0.2truein
{\bf Quantum aspects of gravitation}\\
Radu Roiban, {\it Classical Gravitation from Quantum Scattering Amplitudes: high orders in the post-Minkowskian approximation for binary systems}\\
Ivan Agullo, {\it Quantum bounce as the origin of the anomalies in the CMB}\\
Astrid Eichhorn, {\it Towards implications of asymptotically safe quantum gravity for particle physics}\\
\vskip0.2truein 
{\bf Progress in analytical relativity}\\ 
Gary Horowitz, {\it A Mysterious Cosmic Censorship - Weak Gravity Connection}\\
Justin Vines, {\it TBA}\\
Scott Hughes, {\it Using the large mass-ratio limit to understand the two-body problem in general relativity}\\
\vskip0.2truein
{\bf Tests of general relativity}\\
Cliff Will, {\it Zombie alert! Solar system tests of GR are still alive}\\
Leo Stein, {\it Tests of GR with Gravitational Waves and Numerical Relativity}\\
Dimitrios Psaltis, {\it Tests of General Relativity with black hole shadows}\\
\vskip0.2truein
{\bf Isaacson Award talks and ground based gravitational-wave astronomy}\\  
Bernard Schutz, {\it TBA}\\ 
Bruce Allen, {\it TBA}\\
William East, {\it Probing fundamental physics with gravitational waves}\\ 
\vskip0.25truein
{\bf Nuclear physics with gravitational wave observations}\\
(co-sponsored with DAP)\\
TBA, {\it Status of modeling R-process Nucleosynthesis in BNS and confronting models with observations}\\
Andreas Bauswein, {\it BNS and NS-BH merger simulation in numerical relativity and how they inform nuclear physics}\\
Collin Capano, {\it Neutron Star EoS measurements with GWs from LIGO/Virgo }\\
\vskip0.2truein
{\bf New results from EHT}\\
(co-sponsored with DAP)\\ 
Johnson, {\it TBA}\\
Monika Moscibrodzka, {\it TBA}\\ 
Genzel, {\it TBA}\\
\vskip0.2truein
{\bf NICER and the mass-radius relation of neutron stars}\\
(co-sponsored with DNP)\\
Wynn C. G. Ho, {\it NICER Constraints on Neutron Star Masses and Radii}\\
Anna V. Bilous, {\it NICER Pulse-Profile Modeling Implications for Millisecond Pulsar Physics}\\
Andrew W. Steiner, {\it NICER M-R Implications for the Dense Matter Equation of State}\\
\vskip0.2truein
{\bf Astro2020 Decadal Topics}\\
(co-sponsored with )\\
TBA\\ 
\vskip0.2truein
{\bf What's going on with H0 !?! }\\
(co-sponsored with DAP and GPMFC)\\
TBA\\
\vskip0.2truein
{\bf The history of Cosmology }\\
(co-sponsored with FHP)\\
TBA\\
\vfill\eject

\section*{\centerline {Gravity: New perspectives from strings and higher dimensions}}
\addtocontents{toc}{\protect\medskip}
\addtocontents{toc}{\bf Conference Reports:}
\addcontentsline{toc}{subsubsection}{
\it Gravity: Strings and higher dimensions, by Cynthia Keeler}
\parskip=3pt
\begin{center}
Cynthia Keeler, Arizona State University  
\htmladdnormallink{keelerc-at-asu.edu}
{mailto:keelerc@asu.edu}
\end{center}
The sixth edition of the biennial workshop entitled Gravity: New Perspectives from Strings and Higher Dimensions was held at the Pedro Pascual Centro de Ciencias in Benasque, Spain, from July 14 to July 26 2019.  The workshop brought together experts in general relativity and holography to examine both applications and fundamentals of gravitational physics.

This year's edition of the well-established conference attracted over 90 participants, with expertise in a diverse range of fields including classical general relativity, numerical relativity in both four and higher dimensions, string theory, holography, and quantum information.

Two morning talks left plenty of time for afternoon discussions.  The participants arranged a plethora of organized discussions on a range of topics from field theoretic techniques in the context of gravity to cosmic censorship to black hole thermodynamics.
Below are further highlights on each of the major topics.

\subsection*{Experimental connections}

Three talks focused on gravitational theory as applied to current experimental results.  First, David Mateos gave an overview of holographic techniques for understanding QCD at high densities, with special emphasis on the nonequilibrium physics necessary to accurately model neutron star mergers.  Next, Christian Ecker spoke about holographic models for these neutron star mergers, specifically emphasizing the gravitational wave signals predicted.

In addition to gravitational waves, we also heard about a feature that might be visible in higher resolution versions of the Event Horizon Telescope.  Dan Kapec told us about the numerical relativity results that predict a bright `photon ring' in images of highly spinning black holes, explaining this feature via treating particle motion around near extremal Kerr black holes as an astrophysical example of a critical point.

\subsection*{Field Theoretic Techniques for Gravitational Physics}

Two talks and an afternoon session covered applications of the Sachdev-Ye-Kitaev (SYK) model, a random-matrix conformal field theory.  This model has generated recent interest from the holographic community, since it shows the same quantum chaos signature of black holes in the two-dimensional gravitaton-dilaton theory known as Jackiw-Teitelboim gravity.  In particular, Micha Berkooz presented new exact results in the double-scaled SYK model; this was followed by an afternoon discussion also led by Iosif Bena on wider applications of SYK.  Felix Haehl spoke about the importance of soft (or `reparametrization') modes in maximally chaotic theories at large central charge (including SYK), and in particular about their behavior in larger even-dimensional gravities. Of particular interest for gravitational physicists was the relationship to the chaos physics of Rindler space. 

Eduardo Casali and I also led an afternoon session, reviewing the `classical double copy' which relates solutions of classical gravitational theories to `squares' of classical Yang-Mills theories.  The relationship arises from the color-kinematics duality first found in the study of quantum amplitudes, but recent work has shown its applicability to classical solutions, both exact and perturbative.

\subsection*{Classic Gravitational Problems}

We also discussed progress on several classic gravitational problems. First, Benson Way reviewed the problem of critical collapse and presented a holographic model thereof.  This model reproduces the critical exponent scaling behavior, scale echoing, and naked curvature singularity features of critical collapse (as they appear in asymptotically AdS spacetimes).  Importantly, the (weak) cosmic censorship violation is visible in the divergent stress tensor of the holographic model; it is an open question whether other kinds of cosmic censorship violation, such as the Gregory-Laflamme instability, might have similar holographic signatures.

Oscar Dias discussed the status of \emph{strong} cosmic censorship (CC), or the continuous extendibility of gravitational solutions across Cauchy horizons, in spacetimes with nonzero cosmological constant. In de Sitter spacetimes, Einstein-Maxwell black holes violate strong CC arbitrarily, by tuning close to the extremal charge.  In anti-de Sitter spacetimes, the result depends on dimension; in dimensions four or higher, the Christodolou formulation is still expected to hold (although simple continuity fails).  For $d=3$, however, the BTZ black holes can again be tuned close to extremality, resulting in arbitrarily strong violation of strong CC (which is not removable even by quantum effects).

In addition, Tomas Andrade spoke about numerical results demonstrating weak cosmic censorship violation in sufficiently large dimensions, while Roberto Emparan led an afternoon discussion covering the physics of cosmic censorship violation.

Peter Zimmerman presented holographic results showing asymptotic self-similarity of perturbations in the low temperature limit of AdS black holes, with a universal critical decay exponent.  He connected these results to the manifestation of the Aretakis instability of extremal horizons in holography as a backreaction that causes the decay of the AdS itself.

We also heard updates on the relationship between fluid dynamics and gravity, from Pavel Kovtun and Christiana Pantelidou. Finally, Jiri Podolsky, Vojtech Pravda,  and Alena Pravdova led an afternoon discussion on quadratic gravity.

\subsection*{Entropy}
Once again the relationship between spacetime geometries and entropy was a frequent topic, with two afternoon discussions and three talks.  Max Rota discussed the entropy inequalities, both those which hold for generic quantum systems, and the stronger ones which appear to hold in holographic systems.  Since entropy inequalities are related to geometric areas via the Ryu-Takayanagi principle (and its covariant version, due to Hubeny-Rangamani-Takayanagi; collected referred to as HRRT), geometric relationships between areas in gravitational solutions correspond to relationships between entanglement entropies for holographic field theories. Also in the area of the HRRT principle, I presented a physicist's proof that the 3+1 dimensional metric reconstructible from the full set of entanglement entropies of a 2+1 dimensional holographic field theory must be unique (if it is buildable at all).

On the topic of accounting for black hole entropies via microscopic state counting, Nick Warner reviewed the current understanding of microstate geometries, particularly discussing probes thereof. Jerome Gauntlett spoke about geometric extremization, specifically showing that extremizing the central charge of a class of holographic field theories allows calculation of a supersymmetric index, leading towards a microscopic state counting for black holes with less supersymmetry in AdS$_3$ asymptotics, or with AdS$_2$ asymptotics.  Dario Martelli led an afternoon discussion of entropy in these and other supersymmetric black holes, as studied from CFTs.

Further discussion of black hole thermodynamics was led in an afternoon session by Jennie Traschen, David Kastor, and David Kubiznak, specifically as regards varying the cosmological constant and its relationship to extended black hole thermodynamics.

\subsection*{Flat Holography}

The remaining major topic of the workshop encompassed studies of holography for spaces with flat asymptotics.  Andrea Puhm reviewed recent advances in building a (2-d) field theory dual for spaces with four-dimensional flat asymptotics, highlighting the role played by the flat space asymptotic symmetry group, the soft theorems (particularly the graviton soft theorem), and their relation to the gravitational memory effect.  The current status allows for calculation of correlators in the dual 2-d theory, as well as providing an understanding of its conserved currents and soft theorems. Temple He constructed the relationship between asymptotic symmetries and soft theorems via Ward identities, but for both even and odd dimensions.  

Eliot Hijano presented a connection between AdS/CFT and flat space holography, using a limit intended intended to focus on the `center' of the AdS space.  By considering only interactions within a locally flat causal domain deep inside AdS, it is possible to recover flat space scattering amplitudes, including the infrared divergence of the S-matrix.

\subsection*{Conclusion}
Overall, the broad variety of talks, ample time for further discussion, and beautiful atmosphere of Benasque led to a very successful workshop.  Many of us are looking forward to the next installment in the summer of 2021!

\vfill\eject

\section*{\centerline {The Eighth Meeting on
CPT and Lorentz Symmetry}}
\addtocontents{toc}{\protect\medskip}
\addcontentsline{toc}{subsubsection}{
\it The Eighth Meeting on CPT and Lorentz Symmetry, by Jay Tasson}
\parskip=3pt
\begin{center}
Jay Tasson, Carleton College 
\htmladdnormallink{jtasson-at-carleton.edu}
{mailto:jtasson@carleton.edu}
\end{center}

During the week of May 12th, 2019,
experts from a wide variety of subfields gathered at Indiana University
in Bloomington for the Eighth Meeting on CPT and Lorentz Symmetry (CPT'19).
Like previous events in this triennial series of meetings,
the focus was on tests of these fundamental symmetries
and associated theoretical issues.
As has been the case at the last several events \cite{mog},
there was again significant discussion of local Lorentz invariance,
diffeomorphism invariance,
and gravitational systems.
The meeting consisted of over 50 invited talks spread over 5 days,
a poster session with about 20 posters,
and considerable lively conversation during receptions, meals, and coffee breaks.
A volume,
{\it Proceedings of the Eighth Meeting on CPT and Lorentz Symmetry} \cite{proc}
will be published,
as has been the case for past meetings in the series \cite{procold},
and most of the contributions have already appeared on the arXiv.
This report reviews some of the ideas presented at the meeting
with a focus on gravitational physics.

The search for violations of local Lorentz invariance
continues to be an active area of research.
As a part of the Einstein Equivalence Principle,
local Lorentz invariance is foundational to General Relativity (GR)
and hence must be subject to ongoing scrutiny.
Diffeomorphism invariance is similarly fundamental.
Moreover,
certain suggestions for theories of quantum gravity generate 
violations of these symmetries.
Thus if detected,
violations could provide hints about an underlying theory
that consistently merges GR and quantum physics.
In co-organizer Alan Kosteleck\'y's talk,
we heard some summary and perspective on these foundational ideas \cite{ak}.
We also heard some summative and forward-looking thoughts about the field from
the other co-organizer Ralf Lehnert as high finish to the meeting \cite{rl}.

Over the past several decades,
a framework for systematically testing Lorentz symmetry
called the gravitational Standard-Model Extension (SME)
has been developed \cite{sme}.
The SME is constructed as an effective field-theory expansion
about known physics
that incorporates Lorentz-violating effects at the level of the action.
The meeting included talks that summarized various aspects of the structure
of this framework including a road map for the many complementary limits of SME gravity
that have been explored \cite{jt},
an overview of the structure of the complete theory of linearized gravity \cite{mm},
an introduction to nonlinear gravity in the SME \cite{qb},
and a summary of the structure of the non-gravitational fields in the SME \cite{ba}.
Many of the presentations provided new experimental results,
ideas for new tests,
and theoretical progress
in the context of the general test framework provided by the SME.
The current status of constraints
was summarized in a talk on the {\it Data Tables for Lorentz and CPT Violation} \cite{data},
an annually updated review
of experimental and observational progress.
During the 3 years since the seventh meeting,
the field has again gained access to completely new methods of probing Lorentz symmetry.
Methods that were new at the meeting three years ago have matured,
and long-studied systems are being used in new and increasingly sensitive ways.

It is perhaps no surprise that gravitational waves top the list
of what is new.
The first observation of gravitational and electromagnetic waves
from the same event enabled a vastly improved comparison of their speed.
This in turn offered a 10-order-of-magnitude improvement in the sensitivity
to certain nondispersive Lorentz-violating effects \cite{jt}.
Other Lorentz-violating terms generate dispersion and birefringence in gravitational waves.
The phenomenology of these effects in the context of effective field theory
was discussed \cite{mm,kon}
as were the attempts made to date by
the LIGO Scientific and Virgo Collaborations to seek dispersion \cite{as}.

While gravitational-wave observations continue to provide a powerful new tool for seeking
Lorentz violation in pure gravity,
a powerful new tool for probing Lorentz violation in matter-gravity couplings
has also been realized in space-based Weak-Equivalence Principle tests.
The generality of the SME framework allows for different couplings to Lorentz violation
for different species of matter.
When gravitational couplings are present in the matter sector,
this leads to effective Weak-Equivalence Principle violation.
The flight of the MICROSCOPE mission 
offered another leap forward since the last meeting, and initial limits
achieved from a dedicated analysis of MICROSCOPE data were presented \cite{gm}.
As is often the case,
this dedicated analysis was needed to search for the characteristic frequency dependence of the signals
associated with anisotropic Lorentz violation
that are distinct from those sought in the standard Weak-Equivalence Principle test.

Binary pulsar systems,
long a staple of tests of GR,
continue to deliver novel new results \cite{ls}
as they are spurred on by theoretical developments \cite{qb}.
In this work,
binary pulsar systems were used to place first-ever constraints
on velocity-dependent deviations from the standard $1/r$ dependence of gravity.
Efforts to seek Lorentz violation in laboratory short-range gravity experiments were a new hot topic
at the meeting in 2016 that continue to develop
with planned new experimental designs discussed this year \cite{yfc}.
Another familiar system that bore new fruit
was solar-system tests based on planetary ephemerides \cite{cpl}.
The discussion of Sagnac gyroscopes to test Lorentz violation
matured since the last meeting with presentations on both the full phenomenology \cite{mt}
and the experiments that could do the analysis \cite{adv}.
Developments in atom-interferometer-based inertial sensors were also discussed \cite{ds,yjw},
which have already been used to search for Lorentz violation based on their applications
as gravimeters and Weak-Equivalence Principle tests.
The discussion of interferometric tests continued further with a presentation on the use of
spin-entangled neutrons as tests of spin-gravity couplings \cite{ms},
with potential applications to Lorentz-symmetry tests \cite{zl}.
We heard updates on proposed Weak-Equivalence Principle tests with antimatter \cite{tf,mg}
that could have special implications for the SME,
as well as visually impressive updates on experiments that place neutrons in the quantum states
of the Newtonian potential \cite{ha}
that have implications for CPT and Lorentz symmetry tests \cite{zx}.

Beyond experiments, observational searches,
and the phenomenological work associated with the above tests,
theoretical discussions in a gravitational context were also plentiful.
An understanding of the emergence of local Lorentz and diffeomorphism violation in gravity
via spontaneous or explicit scenarios has been developing
for over a decade.
Riemann-Finsler geometry as a geometric framework for Lorentz-violating theories
has been put forward as a solution to the geometric conflicts that can arise
in the case of explicit Lorentz violation.
This year,
scalar field theories were explored in this context \cite{be}.
There was discussion of scenarios where explicit Lorentz violation
is compatible with Riemannian geometry \cite{rb,yb},
and an illustrative model of spontaneous CPT breaking in the Schwarzschild geometry was presented \cite{dc}.
Specific models that generate terms in the general SME expansion
provide necessary examples,
and noncommutative-gravity theory was discussed in this context \cite{cl}.
A 3+1 decomposition of the SME gravity sector was also presented as a new theoretical tool \cite{nn}.

At this year's meeting,
I was struck,
perhaps even more than in years past,
by the pace of progress in the field.
We've been fortunate in gaining access to novel new ways of probing gravity,
and it's exciting to be a part of a field
where 10-order-of-magnitude improvements are made
between successive triennial meetings.

\begingroup
\renewcommand{\section}[2]{}%

\end{document}